\documentclass[10pt]{article}
\usepackage{graphicx,floatflt,amssymb,epsf}
\usepackage{epsfig,axodraw,citesort}
\def\n1{{\cal{N}}_{1i}} 
\newcommand{\ba}{\begin{array}}    
\newcommand{\ea}{\end{array}}    
\newcommand{\bd}{\begin{displaymath}}    
\newcommand{\ed}{\end{displaymath}}    
\newcommand{\be}{\begin{equation}}    
\newcommand{\ee}{\end{equation}}    
\newcommand{\bea}{\begin{eqnarray}}    
\newcommand{\eea}{\end{eqnarray}}    
\begin{document} 
\begin{flushright} 
LPT Orsay/06-28     
\end{flushright} 

\vskip 5pt 

\begin{center} 
{\Large {\bf Neutrinos in flat extra dimension: towards a realistic scenario}}
\vskip 25pt    
{\bf Asmaa Abada \footnote{E-mail
address: abada@th.u-psud.fr}}, {\bf Paramita Dey \footnote{E-mail
address: Paramita.Dey@th.u-psud.fr}} and 
{\bf Gr\'egory Moreau \footnote{E-mail
address: moreau@th.u-psud.fr}}
\vskip 10pt
$^1${\it Laboratoire de Physique Th\'eorique, Universit\'e
de Paris-sud XI} \\
{\it B\^atiment 210, 91405 Orsay, France} \\

\normalsize 
\end{center} 
\begin{abstract} 
  We study the simple extension of the Standard Model in which an additional
  right handed neutrino propagates along a flat extra dimension, while the
  Standard Model fields are confined on a 3-brane. The fifth dimension is
  $S^1/Z_2$ orbifold compactified. The most generic Lagrangian (thus breaking
  the lepton number) is considered. In this scenario, the neutrino mass can be
  naturally suppressed.  By studying systematically the fundamental parameter
  space, we show that the strong phenomenological constraints on mixing angles
  between active and sterile neutrinos (especially those derived from the SNO
  experiment data) do not conflict with the possibility of generating a
  realistic neutrino mass spectrum.  As a second step, we explore the
  possibility of a successful leptogenesis through the decays of the
  Kaluza-Klein excitations of the right handed neutrino.  \\
  \vskip 5pt \noindent
  \texttt{PACS Nos: 11.10.Kk, 11.30.Fs, 14.60.Pq, 14.60.St, 98.80.Cq} \\
  \texttt{Keywords: Extra Dimensions, Neutrino Physics}
\end{abstract} 

\newpage

\begin{flushright} 
\end{flushright} 

\vskip 30pt

\setcounter{footnote}{0} 
\renewcommand{\thefootnote}{\arabic{footnote}} 

\section{Introduction}
In the recent days, our perspectives in the search for physics beyond
the Standard Model (SM) have enriched considerably with the
developments of theories with extra (compact) dimensions
\cite{extra1,extra2,extra3}. In particular, these extra spatial
dimensions allow to lower the gravity scale down to the TeV
\cite{extra1,extra2}, thus addressing the long standing puzzle of the
gauge hierarchy problem, when gravitational interactions feel these
dimensions. Such TeV-scale Extra Dimensional (ED) models have
attracted a great attention from the experimentalist community, since
one can undertake their precision studies in the next generation
colliders. Depending on the geometry of the extra dimensions (shape
and/or size), and the field localization, a wide variety of extra
dimensional models have been proposed.  In addition to the gauge
hierarchy problem, these models address several other questions in the
different avenues of particle physics, cosmology and string
theory.\newline Among a large number of possible realizations of the
ED models, the scenario, with additional right handed neutrinos (being
gauge singlet, and thus sterile) in flat compact extra dimensions and
SM fields localized on a brane, provides an interesting alternative
\cite{DDG2,ADDM,AP1,IP} capable of accounting for the observed light
neutrino masses \cite{MM}.  Also, the detailed experimental results on
neutrino properties, in the context of this scenario, may be utilized
to shed light on the bulk geometry (for phenomenological studies, see
\cite{DDG2,ADDM,allothers}).
\newline In this paper we study this scenario based on the formalism
of \cite{Bhattacharyya:2002vf}. Our first aim is to determine whether
a realistic neutrino spectrum can still be generated when the
constraints on mixing angles between active and sterile neutrinos are
taken into account, and particularly the severe constraint from the
SNO experiment. Such constraints should indeed apply on the considered
scenario, since the new right handed neutrino, as well as each
component of its infinite massive tower of Kaluza-Klein (KK)
excitations, constitute sterile neutrinos which mix with SM active
neutrinos. We consider the general Lagrangian, which possess two
distinct sources of Lepton number $(L)$ breaking. The phenomenological
constraints on neutrinos were applied, in \cite{DLP}, to the model
with an $L$-conserving Lagrangian assuming no brane shifting.
\newline Our second goal is to address the issue of a successful leptogenesis
within such realistic ED model in agreement with experimental neutrino data.
More precisely, by studying the decays of the heavy KK neutrinos, we explore
the possibilities that these scenarios account for the observed baryonic
asymmetry of the Universe by means of the Fukugita-Yanagida mechanism of
leptogenesis \cite{FY}. In this mechanism an excess of $L$, generated by
out-of-equilibrium $L$-violating decays of the heavy neutrinos (arising with
the $L$-violating terms present in our Lagrangian), is converted into a Baryon
number ($B$) asymmetry through $(B+L)$-violating sphaleron interactions. At
this stage we mention a preliminary work \cite{AP1} where the leptogenesis has
also been considered in the same context of flat extra dimensions, but with an
extended Higgs sector compared to the SM and without considering the various
experimental neutrino data. \newline We will consider the case where gravity
may propagate in an higher dimensional space of $[1+(3+n_g)]$ dimensions,
where $n_g\ge 1$. Hence we have two variants of the scenario, one with $n_g=1$
and the other with $n_g>1$. In all cases, the SM particles are localized on a
$[1+3]$ dimensional subspace ($3$-brane). For simplification reason, the
sterile neutrinos are assumed to propagate in $[1+(3+1)]$ dimensions.
\newline The organization of the paper is as follows. We begin with a brief
outline of the essential features of the extra dimensional scenario in Section
\ref{mhdmn}. The relevant phenomenological constraints on neutrinos are
discussed in Section \ref{ars}, while our numerical results are presented and
discussed in Section \ref{R&D}. Next in Section \ref{seclep}, the issue of
leptogenesis is addressed. Lastly, we summarize our studies, and conclude, in
Section \ref{conclu}.

\section{Minimal Higher Dimensional Model for Neutrinos}
\label{mhdmn}
In this section we describe the minimal ED framework with a
5-dimensional iso-singlet right handed neutrino added to the field
content of the SM \cite{DDG2,Bhattacharyya:2002vf}. Here, we stick to
one generation of SM neutrinos. The fifth flat dimension, along which
propagates the right handed neutrino, is compactified over a $S^1/Z_2$
orbifold. The SM fields are localized on a 3-brane, whereas gravity
propagates in the bulk.
\newline The leptonic field content is,
\begin{equation}
\label{LSM}
L(x)\ =\ \left( \begin{array}{c} \nu_\ell (x) \\ \ell_L (x) \end{array}
\right) ,\qquad
\ell_R (x)\,,
\end{equation}
as in the SM, where $\nu_\ell$, $\ell_L$, $\ell_R$ are 4-dimensional
Weyl spinors, plus the 5-dimensional singlet neutrino,
\begin{equation}
\label{Nu}
N(x,y)\ =\ \left( \begin{array}{c} \xi (x,y) \\ 
\bar{\eta} (x,y) \end{array} \right)\, ,
\end{equation}
where $y$ parameterizes the compact fifth direction, and $\xi$, $\eta$
are the 2-component spinors. The $y$-coordinate is compactified on a
circle of radius $R$ (the periodic boundary condition $N(x,y) =
N(x,y+2\pi R)$ is imposed) on the singlet neutrino
field. Additionally, under the action of $Z_2$ symmetry, the two
2-component spinors may be associated with opposite parities,
\begin{equation}
\label{yparity}
\xi (x,y)\ =\ \xi (x,-y)\,,\qquad  \eta (x,y)\ =\ - \eta (x,-y)\,.
\end{equation}
The SM fields are localized at the brane, which, just for generality,
we may assume to be at $y=a$, instead of at the orbifold fixed point
$y=0$. For this model, an effective Lagrangian respecting the 4-dimensional
Lorentz invariance reads as,
\begin{eqnarray}
\label{Leff} 
{\cal L}_{\rm  eff} & =& \int\limits_0^{2\pi R}\!\! dy\
\bigg\{\,   \bar{N} \Big(   i\gamma^\mu  \partial_\mu\,  +\,  \gamma_5
\partial_y \Big) N\ -\  \frac{1}{2}\,\Big( M N^T  C^{(5)-1} \gamma_5 N\ +\ {\rm
h.c.}    \Big)         \nonumber\\     
&&+\,\delta   (y-a)\,  \bigg[\, \frac{{h}_1}{(M_F)^{1/2}}\, 
L\tilde{\Phi}^*      \xi\,    +\, \frac{{h}_2}{(M_F)^{1/2}}\,  L 
\tilde{\Phi}^*   \eta\ +\   {\rm h.c.}\,\bigg]\ +\ 
\delta (y-a)\, {\cal L}_{\rm SM}\, \bigg\}\, ,\quad
\end{eqnarray}
where $\tilde{\Phi} = i\sigma_2 \Phi^*$ is the hypercharge-conjugate
of the SM Higgs doublet $\Phi$, with hypercharge $Y(\Phi) = 1$, ${\cal
L}_{\rm SM}$ is the SM Lagrangian, $M$ is the Majorana mass for $N$
(we do not specify its scale for the moment), $C^{(5)}$ is the
5-dimensional charge conjugation operator and $M_F$ is the fundamental
$n_g$-dimensional gravity scale, given by
\bea
\label{cutoff}
M_P=(2\pi M_F R)^{n_g/2}M_F,  
\eea 
for the simple case where all the compactification radii are of equal
size $R$, $M_P$ being the 4-dimensional Planck scale. A Dirac mass
term $m_D \bar{N}N$ is not allowed in Eq.(\ref{Leff}) because of the
$Z_2$ discrete symmetry. Note that the Majorana mass term for $N$
breaks the 5-dimensional Lorentz invariance as in
\cite{RamLuk}. \newline Following Eq.(\ref{yparity}), the 2-component
spinors $\xi$ and $\eta$ can be expanded in Fourier series as,
\begin{eqnarray}
\label{xi}
\xi (x,y) &=& \frac{1}{\sqrt{2\pi R}}\ \xi_0 (x)\ +\ 
\frac{1}{\sqrt{\pi R}}\ \sum_{n=1}^\infty\, \xi_n (x)\ 
\cos\bigg(\,\frac{ny}{R}\,\bigg)\,,\\
\label{eta}
\eta (x,y) & =& \frac{1}{\sqrt{\pi R}}\ \sum_{n=1}^\infty\, \eta_n (x)\ 
\sin\bigg(\,\frac{ny}{R}\,\bigg)\,,
\end{eqnarray}
where the chiral spinors $\xi_n(x)$ and $\eta_n(x)$ form an infinite
tower of KK fields.  Using these expansions and integrating out the
$y$-coordinate, the effective Lagrangian reduces to,
\begin{eqnarray}
\label{Leff1KK}
{\cal L}_{\rm eff} & = & {\cal L}_{\rm SM}\ +\ \bar{\xi}_0
( i\bar{\sigma}^\mu \partial_\mu) \xi_0\ 
+\ \Big(\, \bar{h}^{(0)}_1\, L\tilde{\Phi}^* \xi_0\ -\
\frac{1}{2}\, M\, \xi_0\xi_0\ +\ {\rm h.c.}\,\Big)\
 +\ \sum_{n=1}^\infty\, \bigg[\, \bar{\xi}_n 
( i\bar{\sigma}^\mu \partial_\mu) \xi_n\nonumber\\
&& +\, \bar{\eta}_n ( i\bar{\sigma}^\mu \partial_\mu) \eta_n\
+\ \frac{n}{R}\, \Big( \xi_n \eta_n\, +\, \bar{\xi}_n
\bar{\eta}_n\Big) -\ \frac{1}{2}\, M\, 
\Big( \xi_n\xi_n\, +\, \bar{\eta}_n\bar{\eta}_n\
+\ {\rm h.c.}\Big)\nonumber\\
&& +\, \sqrt{2}\, \Big(\, \bar{h}^{(n)}_1\, L\tilde{\Phi}^* \xi_n\ +\
\bar{h}^{(n)}_2\, L\tilde{\Phi}^* \eta_n\ +\ {\rm h.c.}\,\Big)\, \bigg]\, 
\end{eqnarray} 
in a basis in which $M$ is positive, and with:
\begin{eqnarray}
\label{h1n}
\bar{h}^{(n)}_1 &=& \frac{h_1}{(2\pi M_F R)^{1/2}}\ 
\cos \bigg(\,\frac{na}{R}\,\bigg)\ =\ \bigg(\,\frac{M_F}{M_{\rm
P}}\,\bigg)^{1/n_g}\  
h_1 \cos \bigg(\,\frac{na}{R}\,\bigg)\ =\ 
\bar h_1 \cos \bigg(\,\frac{na}{R}\,\bigg)\,,\\
\label{h2n}
\bar{h}^{(n)}_2 &=& \frac{h_2}{(2\pi M_F R)^{1/2}}\ 
\sin \bigg(\,\frac{na}{R}\,\bigg)\ =\ \bigg(\,\frac{M_F}{M_{\rm
P}}\,\bigg)^{1/n_g}\  
h_2 \sin \bigg(\,\frac{na}{R}\,\bigg)\ =\
\bar h_2 \sin \bigg(\,\frac{na}{R}\,\bigg)\, .
\end{eqnarray}
For deriving the last two equalities on the right hand sides of
Eqs.(\ref{h1n})-(\ref{h2n}), we have made use of Eq.(\ref{cutoff}).

Eqs.(\ref{h1n}) and Eq.(\ref{h2n}) tell us that the reduced
4-dimensional Yukawa couplings $\bar{h}^{(n)}_{1,2}$ can be suppressed
by many orders depending on the hierarchy between $M_P$ and $M_F$; for
example, if gravity and the bulk neutrino feel the same number of
extra dimensions, say $n_g=1$, then these couplings are suppressed by
a factor $M_F/M_{\rm P} \sim 10^{-15}$, for $M_F \approx 10$~TeV (see
also \cite{DDG2,ADDM}).

From Eq.(\ref{Nu}) we observe that $\xi$ and $\bar{\eta}$ have the
same lepton number. Thus, the simultaneous presence of the two
operators $L\tilde{\Phi}^*\xi$ and $L\tilde{\Phi}^*\eta$ in
Eq.(\ref{Leff1KK}) leads to lepton number violation. If the brane were
located at one of the two orbifold fixed points ($y=0$ or $\pi R$),
the operator $L\tilde{\Phi}^*\eta$ would have been absent in
Eq.(\ref{Leff1KK}) as a consequence of the discrete $Z_2$
symmetry. The two operators can coexist, leading to breaking of the
lepton number, if we allow the brane to be shifted by an amount $a
(\ne 0)$ from the orbifold fixed points. In fact, it is possible to
perform such a shifting of the brane (even in a continuous way)
respecting the $Z_2$ invariance of the original higher dimensional
Lagrangian under certain restrictions in Type-I string theories
\cite{GP}. As indicated in \cite{DDG2,Bhattacharyya:2002vf}, the $Z_2$
invariance can be taken care of by allowing the replacements,
\begin{eqnarray}
\xi\, \delta (y - a) &\to & \frac{1}{2}\; \xi\,\Big[\, \delta ( y - a)\: +\:
\delta ( y + a - 2\pi R) \, \Big]\,,\nonumber\\
\eta\, \delta (y - a) &\to & \frac{1}{2}\; \eta\,\Big[\, \delta ( y - a)\: -\:
\delta ( y + a - 2\pi R) \, \Big]\,,
\end{eqnarray}
with $0\le a < \pi R$ and $0\le y \le 2\pi R$. Obviously, a
$Z_2$-invariant implementation of brane-shifted couplings would
require the existence of at least two branes placed at $y= a$ and $y =
2\pi R -a$. A remarkable feature of the brane-shifted framework was
pointed out \cite{Bhattacharyya:2002vf}, where it has been shown that
in such a framework it is possible to completely de-correlate the
effective Majorana-neutrino mass $\langle m\rangle$, that is
effectively measured, and the scale of light neutrino masses, as to
have $\langle m\rangle$ within an observable range.

Therefore, the Lagrangian (\ref{Leff1KK}) contains two types of
Majorana neutrino mass term (involving respectively the parameters $M$
and $\bar h_2^{(n)}$) which lead both to a breaking of $L$. This
$L$-breaking is a necessary ingredient of leptogenesis.

Following the notations of reference \cite{DDG2}, we now introduce the weak
basis for the KK Weyl-spinors,
\begin{equation}
\label{xieta}
\chi_{\pm n}\ =\ \frac{1}{\sqrt{2}}\, (\,\xi_n\: \pm\:
\eta_n\,),
\end{equation}
followed by a rearrangement of $\xi_0$ and $\chi^\pm_n$ states, such
that, for a given value of $k$ (say, $k=k_0$) the smallest diagonal
entry of the neutrino mass matrix is
\bea
\label{vareps}
\left|\varepsilon\right| = {\rm min}\, \Big( \left| M - \frac{k_0}{R} \right|
\Big) \leq 1/(2R).  \eea
In this new basis the reordered 4-component Majorana-spinor vector
$\Psi_\nu$ is the following,
\begin{equation}
\label{Psinu}
\Psi^T_\nu \ =\ \left[\, 
\left(\! \begin{array}{c} \chi_{\nu_{\ell}} \\ \bar{\chi}_{\nu_{\ell}}
\end{array}\!\right)\,,\ 
\left(\! \begin{array}{c} \chi_{k_0} \\ \bar{\chi}_{k_0} 
\end{array}\!\right)\,,\
\left(\! \begin{array}{c} \chi_{k_0+1} \\ \bar{\chi}_{k_0+1} 
\end{array}\!\right)\,,\
\left(\! \begin{array}{c} \chi_{k_0-1} \\ \bar{\chi}_{k_0-1} 
\end{array}\!\right)\,,\
\cdots\,,
\left(\! \begin{array}{c} \chi_{k_0+n} \\ \bar{\chi}_{k_0+n} 
\end{array}\!\right)\,,\
\left(\! \begin{array}{c} \chi_{k_0-n} \\ \bar{\chi}_{k_0-n} 
\end{array}\!\right)\,,\
\cdots\ \right]
\end{equation}
while the effective Lagrangian for right handed neutrinos reduces to,
\begin{equation}
\label{Lkinorb}
{\cal L}_{\rm kin}\ =\ \frac{1}{2}\, \bar{\Psi}_\nu\,\Big(\, 
i\!\not\!\partial\ -\ {\cal M}^{\rm KK}_\nu\, \Big)\, \Psi_\nu\,,
\end{equation}
where ${\cal M}^{\rm KK}_\nu$ is the corresponding neutrino mass
matrix given by,
\begin{equation}
\label{Morbshift}
{\cal M}^{\rm KK}_\nu\ =\ \left(\! \begin{array}{ccccccc}
0 & m^{(0)} & m^{(-1)} & m^{(1)} & m^{(-2)} & m^{(2)} & \cdots \\
m^{(0)} & \varepsilon & 0 & 0 & 0 & 0 & \cdots  \\
m^{(-1)} & 0 & \varepsilon - \frac{1}{R} & 0 & 0 & 0 & \cdots \\
m^{(1)} & 0 & 0 & \varepsilon + \frac{1}{R} & 0 & 0 & \cdots \\
m^{(-2)} & 0 & 0 & 0 & \varepsilon - \frac{2}{R} & 0 & \cdots \\
m^{(2)} & 0 & 0 & 0 & 0 & \varepsilon + \frac{2}{R} & \cdots \\
\vdots & \vdots & \vdots & \vdots & \vdots & \vdots & \ddots
\end{array}\!\right)\,.
\end{equation}
The most important consequence of such a rearrangement is that the
mass scale $M$, which we did not specify earlier but which could be as
large as possible, is now replaced by the light mass scale
$\varepsilon$.  The mass entries in the first row and first column of
matrix (\ref{Morbshift}) are given by the relation,
\begin{eqnarray}
\label{mk0}
m^{(n)} &=& \frac{v}{\sqrt{2}}\, \bigg[\, \bar{h}_1\, 
\cos\bigg(\frac{ (n-k_0) a}{R}\bigg)\: +\: \bar{h}_2\, 
\sin\bigg(\frac{ (n-k_0) a}{R}\bigg)\,\bigg]\ =\
m\,\cos\bigg(\frac{na}{R}\, -\, \phi_h\,\bigg)\,,\qquad
\end{eqnarray}
with,
\bea 
\label{initialy}
m &=& \frac{v}{2}\sqrt{\frac{h^2_1 + h^2_2}{\pi M_F R}}, \\
\label{phih}
\phi_h &=& \tan^{-1} \left(\frac{h_2}{h_1}\right) + k_0 \frac{a}{R},  
\eea
where $v$ is the vacuum expectation value of the SM Higgs boson.

We observe that the structure of neutrino mass matrix
(\ref{Morbshift}), originating from a combination of Majorana and
Dirac mass terms, is different from the pure Dirac mass case
considered in the preliminary work \cite{DLP}.

\section {Constraints for a Realistic Spectrum}
\label{ars}
The neutrino spectra is obtained by diagonalizing the mass matrix ${\cal
M}^{\rm KK}_\nu$ of Eq.(\ref{Morbshift}). It has been shown in
\cite{Bhattacharyya:2002vf}, for a one-generation case, that the eigenvalue
equation for a restricted class of cases with $a=\pi R/q$, where $q$ is an
integer greater than 1, is easily tractable analytically. We have, in the
generic case, solved the characteristic eigenvalue equation numerically to
obtain the eigenvalues of ${\cal M}^{\rm KK}_\nu$, sticking to the
one-generation case. We have diagonalized the $N \times N$ matrix of
Eq.(\ref{Morbshift}) for a $N$ satisfying $M_F>M_{\rm max}$ (see end of the
paragraph). We have systematically checked that the $N$ value used (typically
$\sim 500$ for the smallest $R^{-1}$ values to $\sim 10$ for the largest
$R^{-1}$ values) constitutes a dimension at which the lightest eigenvalues
converge and are completely stabilized (with increasing $N$). We have
performed a scan over the whole parameter space of the ED model described
above in order to find the regions in agreement with basic experimental
constraints on neutrinos. The total parameter space consists of $R^{-1}$,
$\varepsilon$, the complex Yukawa couplings $h_1$ and $h_2$, the brane-shift
parameter $a$, and the phase $\phi_h$ defined in Eq.(\ref{phih}). From
Eq.(\ref{cutoff}), the effective cut off scale $M_F$ for a given $R^{-1}$ is
completely fixed for $n_g=1$; while for $n_g>1$, it depends on $n_g$ and the
respective compactification radii (respecting $M_F>M_{\rm max}$, where $M_{\rm
max}$ is the mass of the heaviest neutrino eigenstate considered).

While scanning over the total parameter space described above in order
to search for a realistic neutrino mass spectrum, we have to ensure
several constraints described in the following;
\begin{itemize}
\item There exist phenomenological constraints on the mixing angles
  between active neutrinos and sterile ones. These constraints apply
  on the present model, as the 0-mode and KK excitations of the
  additional right handed neutrino behave exactly as sterile neutrinos
  $-$ the right handed neutrino having no electroweak interactions
  (see e.g. \cite{Moreau}).\\ For instance, in a model with a unique
  sterile neutrino, and assuming for simplicity one lepton flavor, the
  mixing angle $\theta_s$ between the active and sterile neutrino is
  constrained typically by $\tan \theta_s^2 \lesssim 10^{-6}-10^{-1} \
  \mbox{for} \ \Delta m^2 \in [10^{-12}-10^{2}] ~ \mbox{eV}^2$
  ($\Delta m^2$ being the mass eigenvalue squared difference)
  \cite{Vissani} from cosmological and astrophysical considerations
  combined with data from atmospheric, solar (including SNO), reactor
  and short base-line experiments. In the limiting case $\Delta m^2
  \gg \Delta m^2_{sun} \sim 10^{-5}$ eV$^2$, the SNO bound on the
  fraction of oscillating sterile neutrinos coming from the sun
  $\eta_s=\sin^2\theta_s/2$ \cite{Vissani}, becomes approximatively
  \cite{Aliani,Petcov}
\begin{equation}
\eta_s \lesssim 1.2 \times 10^{-1} ~~{\rm at} ~1 \sigma.
\label{eq:limitSNO}
\end{equation} 
Similarly, in our framework, one can also take as an approximation
that the $\chi_{k_0}$ state, whose mass $\varepsilon$ is the first
appearing as a diagonal matrix element in Eq.(\ref{Morbshift}), is the
only sterile neutrino that may not decouple from the active SM
neutrino $\nu_{\ell}$. Indeed, by definition of $\varepsilon$ in
Eq.(\ref{vareps}), the following $\chi_{k_0 \pm n}$ states get larger
and larger diagonal mass matrix elements $\varepsilon \pm n/R$ and
thus decouple more and more from $\nu_{\ell}$, given the symmetric
structure of the mass matrix in Eq.(\ref{Morbshift}).  Therefore,
under the above approximation, the SNO bound on $\theta_s$ apply in
our framework also, where the mixing angle $\theta_s$ is deduced only
from the $2 \times 2$ block matrix of Eq.(\ref{Morbshift}):
\begin{equation}
\tan 2 \theta_s  \approx \frac{2 m^{(0)}}{\varepsilon}. 
\label{eq:angleSYM}
\end{equation}
The above SNO bounds impose a significant upper limit on $\theta_s$
leading to a case comparable to the see-saw mechanism \cite{seesaw}
(see also the third point) so that the smallest eigenvalue $m_{\nu_1}$
and the next-to-smallest eigenvalue $m_{\nu_2}$ read as
\begin{equation}
m_{\nu_1}  \approx \frac{m^{(0)}}{\varepsilon} m^{(0)} , \ \ \ 
m_{\nu_2}  \approx \varepsilon .
\label{eq:seesaw}
\end{equation}
The constraint in Eq.(\ref{eq:limitSNO}) imposes
$m^{(0)}\ll\varepsilon$ ({\it c.f.} Eq.(\ref{eq:angleSYM})), which may
be achieved in the following two cases (see Eq.(\ref{mk0})):
\begin{eqnarray}
\label{condition}
m\ll\varepsilon \,, ~~~{\rm and/or} ~~~
\phi_h \simeq \pi/2 + 2q\pi\,, 
\end{eqnarray} 
where $q$ is an integer.  \\
The previous approximations regarding the implications of the constraints on
sterile neutrino mixing will turn out to be useful in the interpretation of
our results.  In order to quantitatively take into account the effects of the
KK tower of neutrino states when applying the main SNO constraint, we have
imposed the bound: $P(\nu_e \to \nu_s)<0.40$ at 90\% C.L. (coming from the SNO
data \cite{SNO} combined with the Super-Kamiokande results
\cite{SKam}). $P(\nu_e \to \nu_s)$ represents the transition probability
between solar SM electron-neutrino $\nu_e$ and sterile neutrino $\nu_s$, and,
is given by $P(\nu_e \to \nu_s) = 1 - \sum_{\ell=e,\mu,\tau} P(\nu_e \to
\nu_\ell)$ where $P(\nu_e \to \nu_\ell) = \sum_n |V_{e n}|^2 |V_{\ell n}|^2$
\cite{Kuo} with $\ell=e$ in the one active flavor case. The sum here is taken
over the index $n$, which labels the neutrino mass eigenstates, and, $V$
diagonalizes the neutrino mass matrix Eq. (\ref{Morbshift}). The sterile KK
excitations are little affected by matter effects \cite{DLP}. Following the
same numerical approach for $N$ as described in the beginning of this section,
we have checked that for the chosen value of $N$ (corresponding to the maximum
value of $n$ truncating the sum involved in the $P(\nu_e \to \nu_\ell)$
expression), $P(\nu_e \to \nu_s)$ reaches a value which is not significantly
sensitive to further increases of the mass matrix dimension. In other terms, a
satisfactory level of convergence can be reached in the summation of KK
excitations.  In fact, the stabilized values of $P(\nu_e \to \nu_s)$ for our
case turn out to be systematically several orders of magnitude smaller than
the combined SNO-Super-Kamiokande bound mentioned previously, in the final
regions of parameter space which respect all the constraints discussed in this
section.
\item Due to the see-saw structure of the mass matrix, the two
  eigenstates $\nu_1$ and $\nu_2$ are respectively composed mainly by
  the SM neutrino $\nu_{\ell}$ and sterile state $\chi_{k_0}$. Hence
  one can apply the phenomenological bound on $m_{\nu_1}$ derived for
  SM neutrinos, which in the case of one flavor is typically
  \cite{Bell}
\begin{equation}
m_{\nu_1} \sim 10^{(-2~ ,~ 0)} \ \mbox{eV}. 
\label{eq:SMlimit1}
\end{equation}
\item We have also ensured that the mass of the
    next-to-lightest\footnote{In our notations we order the mass
    eigenvalues increasingly, i.e. $m_{\nu_1}\le m_{\nu_2}\le
    \cdots$.} eigenstate $\nu_2$ is larger than the mass of the Higgs
    boson, typically
\begin{equation}
m_{\nu_2} \ge 200 \ \mbox{GeV}, 
\label{eq:SMlimit2}
\end{equation}
for our purpose which deals with its possible lepton-number violating decay
into Higgs and leptons: $\nu_2 \to \ell^\pm\phi^\mp$ (the three-body decay
channel of the $\nu_2$ is $\nu_2 \to \ell^\pm \ell^\mp \nu_1$).
\newline The choice of this condition $m_{\nu_2}>200$ GeV is important
for our study of leptogenesis (in Section \ref{seclep}). We could also
have imposed this constraint on any higher-mass neutrino eigenstate
$m_{\nu_{i}}>200$ GeV with $i>2$, and studied the decay $\nu_{i}\to
\ell^\pm\phi^\mp$. However, any higher state $\nu_i$ ($i>2$), in
addition to its decay channel $\nu_i\to\nu_1\ell^+\ell^-$, possesses
several other decay channels: their decays into lighter eigenstates
$\nu_i\to\nu_j\ell^+\ell^-$ or $\nu_i\to\nu_j\nu_1\nu_1$ with $i > j
\ge 2$. How the presence of these additional decay channels is going
to affect the Boltzmann equations and the washout factors, and thus
the resulting lepton abundance, is a matter of non-trivial calculation
that we will not address here. Such additional decay channels do not
exist for the state $\nu_2$, as the decays $\nu_2\to\nu_2\nu_1\nu_1$
and $\nu_2\to\nu_2\ell^+\ell^-$ are not possible kinematically. Here
we consider the possibility of the ED model to yield a successful
leptogenesis via the $\nu_2$ decay. \\ With this bound $m_{\nu_2} \ge
200$ GeV we are in the case where $\Delta m^2 =
m^2_{\nu_2}-m^2_{\nu_1} \gg \Delta m^2_{sun}$ so that the relevant SNO
bound is the one given in Eq.(\ref{eq:limitSNO}). Furthermore, this
condition forces $m_{\nu_2}$ to be much larger than $m_{\nu_1}$ and so
constitutes another justification for approximating the two lightest
eigenvalues by the see-saw formulas (\ref{eq:seesaw}).
\end{itemize}
At this stage the following remarks on neutrino mass are in
order. There are typically three different mass suppression mechanisms
in the extra dimensional scenario, that may suppress the mass of the
SM neutrino.
\begin{enumerate}
\item The usual see-saw like suppression, from a heavy Majorana mass
  $M$ for the right handed neutrino.
\item An effective ED see-saw type mechanism resulting from the
  massive KK tower \cite{DDG2,AP1}, where the KK excitations of the
  additional right handed neutrino play the r\^ole of the right handed
  neutrino in the usual see-saw mechanism.
\item A suppression of the Yukawa couplings from the ED wavefunction
  overlap mechanism affecting the neutrino mass via the ED volume
  factor $( M_F R)^{1/2}$ (see Eq.(\ref{initialy})) and the factor
  $\cos(na/R-\phi_h)$ (see Eq.(\ref{mk0})).
\end{enumerate}
The main motivation for working in an ED scenario is to generate small
SM neutrino masses using the second and third mechanisms only,
i.e. the usual see-saw mechanism does not constitute the main
suppression origin. Is that possible? As we have already discussed,
the diagonal matrix element $\varepsilon$ contributes dominantly to
the eigenvalue $m_{\nu_1}$ since it is the lightest one. Hence in
order to insure that the neutrino mass suppression is not mainly due
to the usual see-saw mechanism, one can impose the condition $M \neq
\varepsilon$, which is equivalent to $M>1/2R$ (see upper left $2
\times 2$ block matrix in Eq.(\ref{Morbshift})). This condition can
always be satisfied since the mass scale $M$ never appears in this
scenario directly: the presence of the infinite tower of KK states
guarantees that only the value of $M$ modulo $R^{-1}$, namely
$\varepsilon$ (in absolute values) plays a physical r\^ole
\cite{DDG2,Bhattacharyya:2002vf}.

We now discuss the ranges considered for each fundamental free
parameter.  First we made sure that it is mainly the second and/or the
third mechanism (described previously) that suppresses $m_{\nu_1}$, by
letting the numerical values of the Yukawa couplings ${h}_1$ and
${h}_2$ to be within the natural range $\sim 0.1-5$ with the
respective phases ranging between 0 and $\pi$. The phase $\phi_h$ was
also varied within 0 and $\pi$. We varied $R^{-1}$ from a few tens of
eV up to $\sim 10^{19}$ GeV. For $R^{-1} \gtrsim 10^{19}$ GeV, the
heaviest state considered has a mass $M_{\rm max}$ greater than $M_F$,
i.e.  our low energy effective theory looses its predictivity. The
range of $R^{-1}$ being so huge, we considered a logarithmic binning
of the scale of $R^{-1}$.  For each $R^{-1}$, the parameter
$\varepsilon$ may also vary within a large range, especially when
$R^{-1}$ is large, following Eq.(\ref{vareps}). So we also scan over
$\varepsilon$ by choosing randomly the values of $\alpha$ in
$\varepsilon=10^{\alpha}$. The brane shift parameter $a$ was scanned
in a similar way in $0<a<\pi R$. Finally, we performed a scan by {\it
randomly} choosing $\sim 10^6$ points, so that each point corresponds
to a {\it distinct but random} combination of all the parameters.

\section{Results and Discussion}
\label{R&D}
When scanning over the parameter space $\{R^{-1},\varepsilon,\phi_h,
h_1, h_2,a\}$ we have imposed a realistic spectra, i.e. one respecting
all the constraints mentioned in the previous section, by using the
numerical estimation of this neutrino mass spectra. We have explored
the two typical situations allowing for such realistic spectra ({\it
c.f.}  Eq.(\ref{condition})) within two scenarios characterized by
$n_g=1$ and $n_g>1$. Our observations and discussion for both the
scenarios are given below. We have also included, in each case,
possible analytical interpretations of our results.

\subsection{Scenario with $n_g=1$}
\label{ng1}
\begin{enumerate}
\item {\bf The scale $M_F$:}~We recall (see Eq.(\ref{cutoff})) that
  the cut off scale $M_F$ depends only on $R^{-1}$ for $n_g=1$. As we
  increase the scale of $R^{-1}$ from a few tens of eV to $\sim
  10^{19}$ GeV, the corresponding scale $M_F$ increases from $\sim
  10^{9}$ GeV to $\sim 10^{19}$ GeV, following Eq.(\ref{cutoff}). This
  makes the scale $m$ in Eq.(\ref{initialy}) increasing from $\sim
  10^2$ eV to $\sim 10^3$ GeV.
\item {\bf The parameter $R^{-1}$:}~ We observed that the energy scale
  $R^{-1}$ has to be $\gtrsim 400$ GeV (corresponding to $M_F \gtrsim
  10^{13}$ GeV) for generating a realistic neutrino spectrum (see
  Fig.\ref{graph1}).  The reason is that $R^{-1} \ge 2
  \left|\epsilon\right|$ (from Eq.(\ref{vareps})) with $\epsilon
  \approx m_{\nu_2}$ from Eq.(\ref{eq:seesaw}) and we have imposed
  $m_{\nu_2} \ge 200$ GeV. With regard to the neutrino mass, what is
  the physical meaning of this lower bound on $R^{-1}$?  We know from
  Eq.(\ref{cutoff}), that as $R^{-1}$ increases, $M_F$ also increases
  but at a much slower rate compared to $R^{-1}$. As a result, the
  quantity $1/\sqrt{M_F R}$ increases with $R^{-1}$ so that the volume
  suppression mechanism for Yukawa couplings becomes less and less
  effective (see Eqs.(\ref{h1n}), (\ref{h2n})), for a fixed
  $\cos\phi_h$. Besides, the hierarchy between $m^{(0)} \propto
  1/\sqrt{M_F R}$ and $R^{-1}$ increases with $R^{-1}$, so that the
  effective ED see-saw mechanism becomes more effective in the
  reduction of $m_{\nu_1}$ if $\varepsilon$ is set at its maximum
  value $1/2R$. Indeed the denominator of $m_{\nu_1}$
  (Eq.(\ref{eq:seesaw})) clearly increases at a faster rate compared
  to the numerator $m^{(0)}$ - thus making the see-saw formulae more
  effective as $R^{-1}$ increases. Therefore, the fact that there is a
  minimum value for $R^{-1}$ means that the see-saw mechanism plays at
  least a certain r\^ole in the neutrino mass suppression.
\item {\bf The parameter $\varepsilon$:}~ For all $R^{-1}\gtrsim 400$ GeV,
  there exists a minimum value of $\varepsilon$, starting from which we end up
  with a realistic spectrum for all possible higher values of
  $\varepsilon~(\le 1/2R)$. This minimum value is significantly larger than
  $200$ GeV (recall that $\varepsilon \approx m_{\nu_2} \ge 200$ GeV) for
  $1/R\gtrsim 10^8$ GeV, as we see on Fig.\ref{graph1} which shows the plot of
  $\varepsilon$ versus $1/R$. This can be understood as follows. \newline This
  figure results from the general scan, and thus includes both the cases
  $m\ll\varepsilon$ and $\phi_h\sim \pi/2+2q\pi$.  The three straight lines
  correspond to the $\varepsilon$ values obtained analytically from the
  approximate see-saw relation in Eq.(\ref{eq:seesaw}), for $m_{\nu_1} =
  1$~eV, $0.1$~eV and $0.01$~eV, and $\sqrt{h_1^2+h_2^2}=4$ and $\phi_h=0$
  (i.e. $\cos\phi_h=1$). Therefore, the points obtained close to these lines
  correspond to the case where the $m_{\nu_1}$ suppression is mainly due to
  the configuration (1). The regions far below these lines indicate the
  simultaneous occurrence of both configuration (1) and (2) i.e.
  $m\ll\varepsilon$ as well as $\phi_h\sim \pi/2+2q\pi$. Starting from these
  regions and approaching higher $R^{-1}$ values, $m$ increases, making the
  $m_{\nu_1}$ suppression from (1) less effective. Thus if we are to generate
  the same amount of suppression in $m_{\nu_1}$ (or equivalently the same
  smallness of ratio $m^{(0)}/\varepsilon$) for a fixed $\varepsilon$, the
  $m_{\nu_1}$ suppression coming from case (2) must be more effective, i.e.
  $\phi_h$ has to be closer to $\pi/2$. This condition on $\phi_h$ basically
  reduce the allowed range for $\phi_h$, so that the scan misses the allowed
  $\phi_h$ values. This explains the large vacant area in Fig.\ref{graph1}
  below the three lines. Therefore, the easier way to generate the suppression
  in the higher $R^{-1}$ region is by case (1), i.e. by pushing the minimum
  value of $\varepsilon$ to be still higher to account for the increase in
  $m$. Quite naturally therefore, the available parameter region for higher
  $R^{-1}$ decreases, and is concentrated near maximal values of
  $\varepsilon$. Nevertheless, because of choosing to scan over the $\phi_h$
  angle, the particular regions where $a k_0 \ll R$ and $h_1 \ll h_2$ (leading
  to a $\phi_h$ value close to $\pi/2$) have been missed. Anyway such a
  hierarchy between the Yukawa couplings $h_1$,$h_2$ would have no theoretical
  explanation and, in contrast, the interest of the considered ED model is to
  have $h_1 \sim h_2 \sim 1$ while suppressing the neutrino masses mainly from
  the second and/or third mechanism described above (the purely ED
  mechanisms).
  \newline One must realize that the specific points in
Fig.\ref{graph1}, which are such that $\varepsilon \ll 1/R$,
correspond to a value of $M$ extremely close to $k_0/R$ (see
Eq.(\ref{vareps})) and in turn to a situation of fine-tuning between
the three parameters $M$, $k_0$ and $R$.
\begin{figure} 
\vspace{-10pt}
\centerline{\hspace{-3.3mm}
\rotatebox{0}{\epsfxsize=8cm\epsfbox{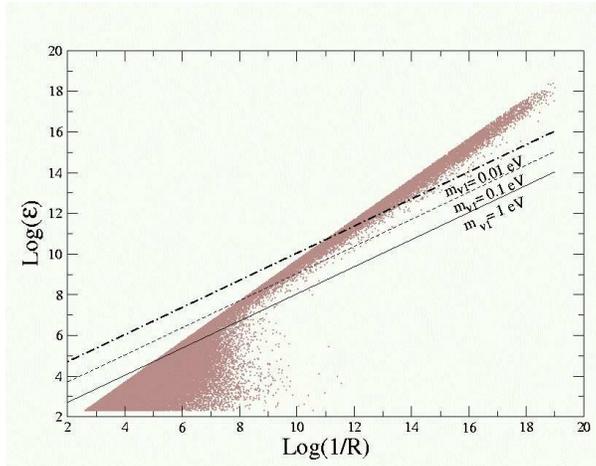}}}
\hspace{3.3cm}\caption[]{ Plot of $\varepsilon$ (in GeV) versus $1/R$ (in GeV)
  obtained from the scan. The three straight lines represent the analytical
  approximation of Eq.(\ref{eq:seesaw}) for the indicated values of
  $m_{\nu_1}$.}  \protect\label{graph1}
\end{figure}
\item {\bf The parameters $\phi_h$ and $a$:}~ First let us discuss the
  spectrum. Since we are always working in a parameter region such that the
  off-diagonal elements of the mass matrix of Eq.(\ref{Morbshift}) are much
  smaller than the diagonal elements, the eigenvalues $m_{\nu_i}$ of the
  matrix, up to a good approximation (see Sec A), are equal to the
  diagonal elements for $i \ge 2$: $m_{\nu_2}\sim\varepsilon$,
  $m_{\nu_3}\sim\varepsilon-1/R$, $m_{\nu_4}\sim\varepsilon+1/R$, and so on.
  In general, the masses of the higher neutrino states can be expressed as
  $m_{\nu_{(2p+1)}}\sim\varepsilon-p/R$,
  $m_{\nu_{(2p+2)}}\sim\varepsilon+p/R$, where $p$ is an integer $\ge 1$. This
  clearly tells us that for a given $R^{-1}$, the nature of the resulting
  spectra is {\it primarily} dictated by the choice of $\varepsilon$. For
  example, if $\varepsilon=1/2R$, the resulting spectrum becomes almost
  pairwise degenerate with $\vert m_{\nu_2}\vert\sim \vert m_{\nu_3}\vert$
  (see Eq.(\ref{eq:seesaw})), $\vert m_{\nu_4}\vert\sim \vert m_{\nu_5}\vert$,
  and so on, while, for all the other choices of $\varepsilon$, the resulting
  spectrum is hierarchical with
  $m_{\nu_2}<m_{\nu_3}<m_{\nu_4}<m_{\nu_5}<\cdots$ \newline In particular, we
  found that for a given $R^{-1}$, the overall nature of the resulting
  spectrum {\it feebly} depends on the choice of the parameters $\phi_h$ and
  $a$. This is expected since these parameters appear {\it only} in the
  non-diagonal mass-matrix elements which are much smaller than the diagonal
  ones for any $\phi_h$ and $a$ values (as those enter via arguments of a
  cosine function). We have found that for any $R^{-1}\gtrsim 400$ GeV, if
  $\varepsilon\sim 1/2R$, then {\it all} possible combinations of $\phi_h$ and
  $a$ yield realistic spectra. Naturally, these possible combinations of
  $\phi_h$ and $a$ include $a=0$ as well. For possible lower values of
  $\varepsilon$, which may not correspond to $m\ll\varepsilon$, a realistic
  spectra requires the second scenario $\phi_h\sim \pi/2+2q\pi$ (as already
  discussed). Note that for this case also, $a=0$ is a natural
  possibility. Thus the important conclusion that we made from these
  observations was that a non-zero brane-shift is {\it not essential} for a
  realistic mass spectrum.
 \begin{figure} 
\vspace{-10pt}
\centerline{\hspace{-3.3mm}
\rotatebox{0}{\epsfxsize=8cm\epsfbox{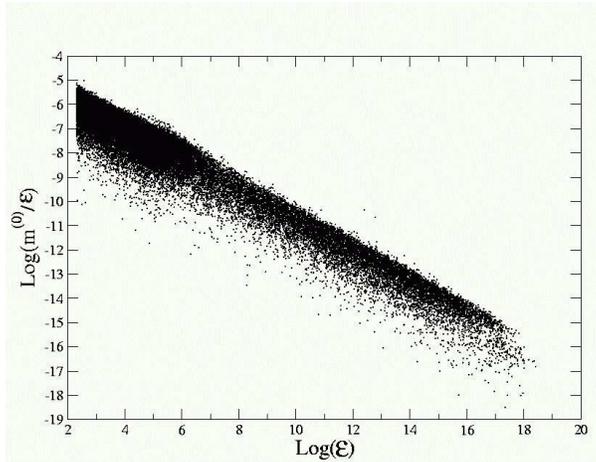}}}
\hspace{3.3cm}\caption[]{ Plot of $m^{(0)}/\varepsilon$ versus $\varepsilon$
  (in GeV).}  \protect\label{graph2}
\end{figure}
  
  Let us compare our observations with those of
  Ref.\cite{Bhattacharyya:2002vf}. The authors of reference
  \cite{Bhattacharyya:2002vf} found that when $\varepsilon=0$, $\phi_h=0$ and
  $a=0$, the mass spectra consists of massive pairwise degenerate neutrino
  states without any massless state. They also found that when
  $\varepsilon=1/2R$, $\phi_h=0$ and $a=0$, there is an almost massless state,
  and the remaining spectra consists of pairwise degenerate massive neutrino
  states. We have found the same observation for these two points.\\ They
  further found that when $\varepsilon\ne 0$ and $\ne 1/2R$, $\phi_h=0$ and
  $a=0$, there is no massless state and the mass spectra is hierarchical. We
  indeed find an {\it almost massless} neutrino state in this case too, the
  overall spectrum being hierarchical.  \\ They also found that {\it unless}
  $\varepsilon=1/2R$, $\phi_h=\pi/4$ and $a=\pi R/2$, the mass spectrum
  consists of massive non-degenerate KK neutrinos. We find indeed that for
  $\varepsilon=1/2R$, $\phi_h=\pi/4$ and $a=\pi R/2$, the high neutrinos
  states are pairwise degenerate. \newline However, we found that for several
  other combinations of these three parameters one can also get a realistic
  mass spectrum. For example, we mentioned before that for $\varepsilon=1/2R$,
  we always get the spectra consisting of an almost massless state and other
  states pairwise degenerate, irrespective of the numerical values of $\phi_h$
  and $a$. Our larger allowed domain of parameter space is due to the
  additional de-correlation condition required in \cite{Bhattacharyya:2002vf},
  and not considered here. In particular, a non-zero brane-shift parameter is
  found to be necessary in \cite{Bhattacharyya:2002vf} in order to account for
  this complete de-correlation of the effective Majorana mass term $\langle
  m\rangle$ and the scale of light neutrino mass (the motivation being to have
  $\langle m\rangle$ in a potentially observable range). In contrast, we find
  the brane-shift to be not essential for a realistic mass spectrum.
\item {\bf The quantity $m^{(0)}$:}~ We finish this section by
  determining which of the two possible ED mechanisms responsible for
  the suppression of $m_{\nu_1}$ (discussed in Sec\ref{ars}) plays the
  dominant r\^ole (for different values of $R^{-1}$). Our discussion
  is based on the plot of $m^{(0)}/\varepsilon$ versus $\varepsilon$
  in Fig.\ref{graph2}. One first observes clearly from the figure that
  systematically $m^{(0)} \le 10^{-5} \varepsilon$ which means that
  the effective ED see-saw mechanism plays {\it always} a significant
  r\^ole in reducing $m_{\nu_1}$ (see Eq.(\ref{eq:seesaw})), as we
  have already noticed. One see typically that the $m_{\nu_1}$
  suppression, relative to the electroweak symmetry breaking scale ($v
  \sim 10^{2}$ GeV), can be due at $50\%$ from the effective ED
  see-saw mechanism and at $50\%$ from the higher dimensional
  mechanism based on wave function overlap, a situation corresponding
  to $\varepsilon \sim 10^{2}$ GeV and $m^{(0)} \sim 10^{-3}$ GeV
  (such that $m^{(0)}/v\sim m^{(0)}/\varepsilon \sim 10^{-5}$). The
  suppression can even come {\it purely} from the effective ED see-saw
  mechanism, namely $m^{(0)}/\varepsilon\sim 10^{-11}$ with
  $m^{(0)}\sim v\sim 10^2$~GeV. Thus the fact that
  $m^{(0)}/\varepsilon\lesssim 10^{-5}$ means that the ED see-saw
  mechanism is systematically the dominant one.
  
  This can interpreted in the following terms. All the conditions that we have
  imposed resulted in a lower bound on $R^{-1}$ (see above)\footnote{In
  particular, the SNO bound constrains the mixing angles between active and
  sterile neutrinos, by this way forcing typically the KK masses ($\propto
  1/R$) to increase.}. Now, as $R^{-1}$ increases, the extra compact bulk
  space decreases, so that the wavefunction overlap factor increases, making
  the suppression of $m_{\nu_1}$ from the geometrical mechanism less
  effective. Hence, the different constraints (including the SNO one) limit
  the wave function overlap mechanism effect. Nevertheless, thanks to the ED
  see-saw mechanism, a realistic spectrum can still be generated.
\end{enumerate}
\subsection{Results for the $n_g>1$ scenario} 
\label{reg5}
We consider now another scenario where the right handed neutrino
propagates only along one extra dimension whereas gravity propagates
in the whole bulk, the number $n_g$ and sizes of extra dimensions
being adjusted (with Eq.(\ref{Morbshift}) when all of them have the
same size $2\pi R$) such that $M_F$ is as low as $\sim {\cal O}(1)$
TeV in order to address the gauge hierarchy problem. Recall that the
solution to the gauge hierarchy problem within the ADD model for
$n_g=1$ is excluded by experimental arguments relative to the Newton
law modification\footnote{In our case for the $n_g=1$ scenario, the
cutoff scale was set around $10^{13}$ GeV and more. This model, for
sure, did not address the gauge hierarchy issue.}.

The radius, as found previously, is bounded from above by $R^{-1}
\gtrsim 400$ GeV, since $R^{-1} \ge 2 \left|\varepsilon\right|$,
$\varepsilon \approx m_{\nu_2}$ and $m_{\nu_2} \ge 200$ GeV. As a
consequence, $m^{(0)} \sim v/\sqrt{M_F R} \gtrsim 10^{2}$ GeV
\footnote{Unless of course $\phi_h\sim \pi/2+2q\pi$, but in order to
decrease significantly the order of magnitude of this lower bound on
$m_{(0)}$, $\phi_h$ should be extremely close to $\pi/2+2q\pi$ which
has a priori no theoretical justification.} as $M_F \sim {\cal O}(1)$
TeV, so that there is no significant suppression in $m^{(0)}$ (and
hence in $m_{\nu_1}$), relative to the electroweak scale $v$, coming
from the ED overlap mechanism. Thus the neutrino mass suppression must
originate principally from the ED see-saw mechanism. For that purpose,
one should have $m^{(0)}/\varepsilon \sim 10^{-11}$ (see
Eq.(\ref{eq:seesaw})) which together with $m^{(0)} \sim v/\sqrt{M_F
R}$ and $\left|\varepsilon\right|<1/2R$ leads to $R^{-1} \gtrsim
10^{26}$ GeV: an amount much larger than the fundamental scale $M_F
\sim {\cal O}(1)$ TeV. Thus, for the $n_g>1$ scenario, generating a
realistic neutrino spectrum and addressing the gauge hierarchy issue
simultaneously, is only possible if there exists a compactification
scale $1/R$ extremely different from the fundamental scale $M_F$
(hence another hierarchy has to be introduced). Besides, such a
scenario is automatically in a non-predictable regime, since the
condition $M_{\rm max}<M_F$ translates into $1/R \lesssim {\rm TeV}$.
\section{Leptogenesis}
\label{seclep}
One interesting question to address is the possibility of having
successful leptogenesis in these ED scenarios. As has been argued in
\cite{AP1}, the infinite KK tower of neutrinos associated with the
5-dimensional sterile neutrino may act as an infinite series of
CP-violating resonators. Thus, we probed the ED scenarios to see
whether they can generate sufficient lepton asymmetry.
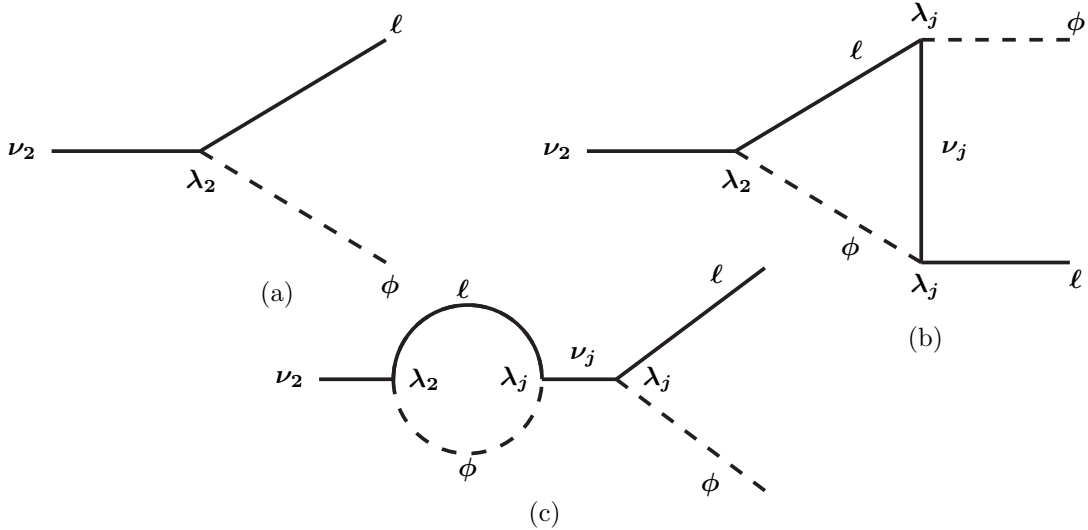
\begin{figure}[t]
\unitlength 1mm
\SetScale{2.8}
\begin{boldmath}
\begin{center}
\begin{picture}(60,30)(0,-10)
\Line(0,0)(20,0)
\Line(45,15)(20,0)
\DashLine(45,-15)(20,0){2}
\Text(-2,0)[r]{$\nu_2$}
\Text(22,-4)[r]{$\lambda_2$}
\Text(45,17)[l]{$\ell$}
\Text(45,-18)[c]{$\phi$}
\Text(30,-19)[c]{(a)}
\end{picture}
\hspace{1cm}\begin{picture}(60,30)(0,-10)
\Line(0,0)(20,0)
\Line(45,15)(20,0)
\DashLine(45,-15)(20,0){2}
\Line(45,15)(45,-15)
\DashLine(45,15)(65,15){2}
\Line(45,-15)(65,-15)
\Text(-2,0)[r]{$\nu_2$}
\Text(22,-4)[r]{$\lambda_2$}
\Text(35,13.5)[l]{$\ell$}
\Text(35,-13)[c]{$\phi$}
\Text(45,-18)[c]{$\lambda_{j}$}
\Text(45,18)[c]{$\lambda_j$}
\Text(45,-25)[c]{(b)}
\Text(49,0)[c]{$\nu_j$}
\Text(65,17)[c]{$\phi$}
\Text(65,-17)[c]{$\ell$}
\end{picture}\hspace{1cm}
\begin{picture}(60,30)(0,-10)
\Line(0,0)(10,0)
\CArc(20,0)(10,0,180)
\DashCArc(20,0)(10,-180,180){2}
\Line(30,0)(40,0)
\Line(60,15)(40,0)
\DashLine(60,-15)(40,0){2}
\Text(-2,0)[r]{$\nu_2$}
\Text(16,0)[r]{$\lambda_2$}
\Text(28,0)[r]{$\lambda_j$}
\Text(52,14)[l]{$\ell$}
\Text(52,-14)[c]{$\phi$}
\Text(47,0)[r]{$\lambda_j$}
\Text(37,3)[r]{$\nu_j$}
\Text(21,-12)[r]{$\phi$}
\Text(20,12)[r]{$\ell$}
\Text(30,-18)[c]{(c)}
\end{picture}
\end{center}
\end{boldmath}\vskip 0.3cm
\caption{Diagrams for the CP-asymmetry decays, at tree (a) and loop (b)-(c)
level, of the second neutrino eigenstate $\nu_2$ into the charged lepton and
the charged Higgs boson.}  \protect\label{lepto}
\end{figure}
\subsection{Additional Constraints and Expectations}
\label{secace}
In the previous section we saw that the lightest neutrino $\nu_1$ in
the generated spectrum fulfills the absolute scale bound.  We also
constrained the next to lightest neutrino $\nu_2$ to be heavier than
the Higgs mass, so that it may decay at a temperature corresponding to
its mass and produce lepton asymmetry \cite{FY}. The out-of
equilibrium condition may be written as the following condition on the
decay width of ${\nu_{2}}$,
\bea \Gamma_{\nu_{2}} \simeq \frac{K_{_{22}} {m_{\nu_2}}}{8\pi} <
{H(T=m_{\nu_{2}})},\quad K_{_{jj}}\equiv [\lambda^\dag\lambda]_{_{jj}},
\label{gamma} 
\eea 
where $H$ is the Hubble expansion rate at $T=m_{\nu_{2}}$, and
$\lambda_j$ is the modified Yukawa coupling for the neutrino mass
eigenstate $\nu_j$. The initial Yukawa couplings, $\lambda'$, can be
read from Eq.(\ref{mk0}): $m^{(j)} = \lambda'_{j} v/\sqrt{2}$. When
the complex symmetric mass matrix in Eq.(\ref{Morbshift}) is
diagonalized, the weak states $\Psi_\nu$ of Eq.(\ref{Psinu}) transform
to the mass eigenstates $\Psi^{\rm mass}_\nu$ following $\Psi_\nu=V
\Psi_\nu^{\rm mass}$, where $V$ is the diagonalizing matrix.  Thus the
initial Yukawa couplings for right handed neutrinos $\lambda'_{i}$
also transform into the modified Yukawa couplings $\lambda_{j}$ via
the $V$ matrix.  We note the Yukawa couplings $\lambda_i$ instead of
$\lambda_{i1}$, where the index $1$ is fixed as it corresponds to the
unique left handed SM lepton (we remind that we work under the
simplification assumption of one family) which moreover does not
develop a KK tower.  The condition in Eq.(\ref{gamma}) is the first
filter to ensure a successful thermal leptogenesis.  The other
necessary condition is to generate enough CP asymmetry, the asymmetry
being defined by,
\bea 
\epsilon_{_{\rm CP}}={\Gamma(\nu_2\to \ell \phi)-\Gamma(\nu_2\to \bar \ell
  \bar\phi)\over \Gamma(\nu_2\to \ell \phi)+\Gamma(\nu_2\to \bar \ell
  \bar\phi)}.
\label{epcp}
\eea
This asymmetry gets a non-zero value due to the interference between
the tree level diagram of Fig.\ref{lepto}a and one-loop diagrams of
Fig.\ref{lepto}b and \ref{lepto}c.  The asymmetry is given by,
\bea
\epsilon_{_{\rm CP}} = \frac{1}{8\pi K_{_{22}}} \sum_{j\neq 2} {\mathrm Im}
  [K_{_{2j}}^2] ~ f(m_{\nu_j}^2/m_{\nu_2}^{2}),
\label{eps1} 
\eea
where $f$ is the loop factor \cite{roulet}, given by  
\bea 
f(x) &=& \sqrt x \left[1 - (1+x)\ln\frac{1+x}{x} + \frac{1}{1-x}
\right], 
\label{floop}
\eea
in the case of a hierarchical neutrino spectrum. Here $\nu_2$, the
next-to-lightest neutrino, is the external (i.e. decaying) mode and
$\nu_j$ the exchanged eigenstate (see Fig.\ref{lepto}). The amount of
CP asymmetry generated can be magnified by the loop function and/or
the Yukawa couplings.  However, due to the large hierarchy between the
lightest first state $\nu_1$ and the next-to-lightest state $\nu_2$,
much enhancement from the Yukawa couplings cannot be expected, as we
will discuss more precisely in the next subsection. The other
enhancement may then come from the loop function via resonances. The
discussion on the characteristics of the generated KK neutrino
spectrum in the previous section clearly shows that we are indeed
often in this favorable resonance situation ($\Leftrightarrow
\varepsilon \sim 1/2R$), since the neutrino eigenstates are almost
pairwise degenerate over a wide range of the obtained parameter
space. In this situation, the self energy contribution to the loop
factor, which is the last term in Eq.(\ref{floop}), is modified to
\cite{PU03}
\bea 
f^{\rm res} (m_{\nu_j}^2/m_{\nu_2}^{2})
&=&\frac{(m_{\nu_2}^2-m_{\nu_j}^2)m_{\nu_2}
\Gamma_{\nu_j}}{(m_{\nu_2}^2-m_{\nu_j}^2)^2+m_{\nu_2}^2 \Gamma_{\nu_j}^2},
\eea
where $\Gamma_{\nu_j} \simeq {K_{jj}m_{\nu_j}}/{8\pi}$ is the decay width of
state $\nu_j$.

\subsection{Results and Discussion}
As before, we have scanned the total parameter space with the
additional constraints mentioned above concerning leptogenesis,
concentrating on the scenario with $n_g=1$. We checked that within the
previously obtained parameter space, the first test to ensure
leptogenesis (represented by Eq.(\ref{gamma})), is satisfied.
\newline As shows Eq.(\ref{eps1}), the source of a non-zero CP
asymmetry is the difference of phases of the Yukawa couplings. Since
we are working with one generation of neutrinos, there is a single
(common) phase associated with the Yukawa couplings $\lambda'$ (the
one associated with the SM neutrino). However, the final (modified)
Yukawa couplings $\lambda_j$ may still be different for different
neutrino mass eigenstates $\nu_j$, as $V$ itself is a complex
matrix. The resulting difference of phases, generated in the process
of diagonalization of the mass matrix as described above, leads to a
non-zero value of ${\mathrm Im}{[\lambda^\dag\lambda]^2_{2j}}$.
\newline However, although non-zero, ${\mathrm
Im}{[\lambda^\dag\lambda]^2_{2j}}$ is small due to the large hierarchies among
the elements of the mass matrix, which in turn is essential to generate a
realistic neutrino spectrum with small active-sterile mixing angles. We find
numerically that the resonance enhancement in $\epsilon_{_{\rm CP}}$ from the
loop function cannot compensate this suppression from the Yukawa couplings in
order to ultimately generate a sufficient CP asymmetry for a successful
leptogenesis: $\epsilon_{_{\rm CP}}>10^{-8}$ \cite{lept} (assuming the usual
orders of magnitude for the washout factors).  Thus, although the pairwise
quasi-degenerate KK neutrino spectrum for one-generation of SM neutrino is
favorable for yielding a good value of $\epsilon_{_{\rm CP}}$, but the
suppression of the same from the Yukawa couplings turns out to be too
dominant.

A possible alternative for a potentially successful leptogenesis could
be to probe a more realistic model with three generations of SM
neutrinos on the 3-brane, with three extra right handed neutrinos in
the bulk.  The mass matrix in Eq.(\ref{Morbshift}) in that case would
have a similar structure with $h_1$ and $h_2$ replaced by $3\times 3$
matrices in the flavor space. The masses of the $n^{\rm th}$ neutrino
states would remain approximately the same for each generation and the
differences in the masses of the three light neutrinos would only come
from the Yukawa couplings. This initial difference in the Yukawa
couplings could be enhanced further during the process of
diagonalization of the mass matrix. This is in contrast to the
one-generation case where we had to start from a single Yukawa
coupling, and the process of matrix diagonalization was the {\it only}
source to create a difference in the final Yukawa coupling
phases. Therefore, we could expect a much larger contribution to the
CP asymmetry from the ${\mathrm Im}{[\lambda^\dag\lambda]}$ factor,
compared to the one-generation case.

\section{Conclusions}
\label{conclu}
In this work, we have considered an $L$-breaking extra dimensional
scenario, with the SM particles localized on a 3-brane and an
additional right handed neutrino propagating along an extra dimension,
providing new mechanisms for suppressing the neutrino mass.  We probed
the possibilities of this model to account simultaneously for a
realistic neutrino mass spectrum and a sufficient lepton asymmetry
(from the decays of the KK neutrinos). We considered two variants of
the above scenario.

The first is when this extra dimension is the only available one
($n_g=1$).  Then there exist fundamental parameters which give rise to
a neutrino spectrum respecting the experimental constraints, on
neutrino masses and active-sterile neutrino mixings (like the SNO
bounds). We have found that in this framework, the neutrino mass
suppression originates dominantly from the ED see-saw mechanism and
partially from the ED wave function overlap effect.  In fact, the SNO
data limit the effect of the wave function overlap but a sufficiently
small neutrino mass can be generated thanks to the ED see-saw
mechanism.  For certain parameter domains, the spectra in this
scenario consists of pairwise quasi-degenerate heavy (KK) neutrinos,
together with a light neutrino. We observed that under the constraints
on neutrino masses that we have imposed, a ``brane shifted'' framework
of the extra dimensional model is not essential for generating a
realistic mass spectra. Also, the inverse radius of compactification
has to be pushed to $\sim 400$ GeV (or more).  This in turn results in
very tiny phase differences among the modified Yukawa couplings (after
the process of mass matrix diagonalization) so that it is not possible
to create a CP asymmetry large enough to insure a successful
leptogenesis.  Indeed, such a tininess is not sufficiently compensated
by the enhancement from the resonant loop function (related to the
decay of the KK neutrinos) which is due to the pairwise
quasi-degeneracy property of the spectrum.

In the second variant where gravity can propagate in more extra
dimensions ($n_g>1$), the fundamental gravity scale is reduced down to
the TeV scale so that the gauge hierarchy problem is addressed through
the ADD approach. We have found that in this framework, the neutrino
mass suppression originates also primarily from the ED see-saw
mechanism and the compactification scale is of an order of magnitude
much larger than the fundamental gravity scale.

\vskip 10pt

\centerline{\bf{Acknowledgments}} We thank R.~Singh for important
clarifications and suggestions regarding several numerical aspects at
different stages of this work. We are also grateful to
G.~Bhattacharyya for reading the manuscript and proposing important
modifications. A.~A. acknowledges support from the ANR-05-JCJC-0023-02
(project NEUPAC). A.~A. and G.~M. acknowledge support from the
ANR-05-BLAN-0163-01 (project Phys@Col@Cos). P.~D. thanks for the grant
from the University of Paris-XI provided during the fulfillment of
this project.

\vskip 20pt
{\Large{\bf A: Approximate analytical estimation of KK-neutrino masses}} 
\label{ap1}

Here we intend to show analytically that when the eigenvalues
($\mathcal{E}_n$ with $n \geq 2$) of mass matrix (\ref{Morbshift}) are
taken equal to the diagonal elements, their correction are much
smaller than the eigenvalues themselves. For that purpose, let us
start with the $4\times4$ form of Eq.(\ref{Morbshift}),
\begin{equation}
\label{mexp1}
{\cal M} =\ \left(\! \begin{array}{cccc}
0 & m^{(0)} & m^{(-1)} & m^{(1)} \\
m^{(0)} & \varepsilon & 0 & 0 \\
m^{(-1)} & 0 & \varepsilon - \frac{1}{R} & 0 \\
m^{(1)} & 0 & 0 & \varepsilon + \frac{1}{R} \\
\end{array}\!\right)\,.
\end{equation}
All the obtained sets of parameters (see discussion on
Fig.\ref{graph1}) lead to a significant hierarchy in $m/\varepsilon$
(case (1) of Eq.(\ref{condition})). Hence the $m^{(n)}$ values are
systematically much smaller than $\varepsilon$, from Eq.(\ref{mk0}),
namely: $m^{(n)}\ll\varepsilon$. We can thus rewrite matrix in
Eq.(\ref{mexp1}) as ${\cal M}={\cal M}_D + {\cal M}_I$, where ${\cal
M}_D$ is the diagonal matrix constructed from ${\cal M}$, and ${\cal
M}_I$ is a perturbation matrix with no diagonal entries given as,
\begin{equation}
\label{mexp2}
{\cal M}_I =\ \left(\! \begin{array}{cccc}
0 & m^{(0)} & m^{(-1)} & m^{(1)} \\
m^{(0)} & 0 & 0 & 0 \\
m^{(-1)} & 0 & 0 & 0 \\
m^{(1)} & 0 & 0 & 0 \\
\end{array}\!\right)\,.
\end{equation}
The zeroth order eigenvalues $\mathcal{E}_i$ of ${\cal M}$ can be
written as, $\mathcal{E}_1 = 0, \mathcal{E}_2 = \varepsilon,
\mathcal{E}_3 = \varepsilon-1/R, \mathcal{E}_4 =
\varepsilon+1/R$. Since ${\cal M}_I$ has all diagonal elements equal
to zero, the lowest order correction to the eigenvalues reads as,
\begin{equation}
\label{mexp4}
\delta\mathcal{E}_n=\sum_{n'\ne n}\frac{(<\psi_n\left|{\cal
    M}_I\right|\psi_{n'}>)^2}{\mathcal{E}_n-\mathcal{E}_{n'}}, 
\end{equation}
where the $\psi$'s are the zeroth order eigenvectors (considering the
non-degenerate case). Thus, for our case we have from the above
relations,
\begin{eqnarray}
\label{mexp51}
\delta\mathcal{E}_1 &=& \frac{(m^{(0)})^2}{-\varepsilon} +
\frac{(m^{(1)})^2}{-\varepsilon-1/R} +
\frac{(m^{(-1)})^2}{-\varepsilon+1/R}\\
\label{mexp52} 
\\ 
\delta\mathcal{E}_n &=& (m^{(n)})^2/\mathcal{E}_n,
\end{eqnarray}
where $m^{(n)}$ is given by Eq.(\ref{mk0}) and $n>1$.  Since
$m\ll\varepsilon$, the first two of the above relations lead to
\begin{eqnarray}
\label{mexp512}
\delta\mathcal{E}_1 &\le&
  m^2\left[\frac{2\varepsilon}{(1/R)^2-(\varepsilon)^2} -
  \frac{1}{\varepsilon}\right], \\ \delta\mathcal{E}_2 &\le& m^2/\varepsilon .
\end{eqnarray}
These inequalities can immediately be generalized to the $N\times N$
mass-matrix as,
\begin{eqnarray}
\label{mexp61}
\delta\mathcal{E}_1 &\le&
m^2\left[\sum_{n=1}^{(N-1)/2}
  \frac{2\varepsilon}{(n/R)^2-(\varepsilon)^2}-\frac{1}{\varepsilon}\right], 
\\ 
\label{mexp62}
\delta\mathcal{E}_2 &\le& m^2/\varepsilon.
\end{eqnarray}
For the maximum value of $\varepsilon~(=1/2R)$, the first term on the
right hand side of Eq.(\ref{mexp61}) is $\sim 2R$ for large $N$, so
that $\delta\mathcal{E}_1\sim 0$. For smaller values of $\varepsilon$,
$\delta\mathcal{E}_1$ would be dominated by the second term on right
hand side. For all the other eigenvalues with $n > 2$, one can derive
$\mathcal{E}_n\ge\varepsilon$.  Thus, in general the numerical value
of $\delta\mathcal{E}_n$ for {\it any} $n$ verifies
\begin{eqnarray}
\label{mexp7}
\left|\delta\mathcal{E}_n\right| \le m^2/\varepsilon.
\end{eqnarray}
So let us study the $m^2/\varepsilon$ value. For $R^{-1} \lesssim
10^{8}$ GeV, $\varepsilon \gtrsim 200$ GeV (see Fig.\ref{graph1}) and
$m \lesssim 10^{-1}$ GeV ({\it c.f.} Eq.(\ref{initialy}) and
Eq.(\ref{cutoff}) in case of the scenario with $n_g=1$).  Hence, one
has $m^2 / \varepsilon \lesssim 10^{-5}$ GeV. Now for $R^{-1} \gtrsim
10^{8}$ GeV, $\varepsilon \sim R^{-1}/2$ and $m \sim 10^{-4} \
(R^{-1}/\mbox{1GeV})^{1/3}$ GeV, so that $m^2 / \varepsilon \sim
10^{-8} \ (R^{-1}/\mbox{1GeV})^{-1/3}$ GeV and thus $m^2 / \varepsilon
\lesssim 10^{-11}$ GeV. Therefore, from Eq.(\ref{mexp7}), we conclude
that one has systematically $\left|\delta\mathcal{E}_n\right| <<
\mathcal{E}_n$ for each $n \geq 2$, since $\mathcal{E}_{n \geq 2}
\gtrsim \varepsilon \gtrsim 200$ GeV.  A similar analysis could be
performed for the case of the degenerate mass spectra (then we would
have to start from the corresponding relation of Eq.(\ref{mexp4}) for
degenerate eigenvalues).

\end{document}